# Monolingual Probabilistic Programming Using Generalized Coroutines

**Oleg Kiselyov**
FNMOC
Monterey, CA 93943

**Chung-chieh Shan**
Rutgers University
Piscataway, NJ 08854

## Abstract

Probabilistic programming languages and modeling toolkits are two modular ways to build and reuse stochastic models and inference procedures. Combining strengths of both, we express models and inference as generalized coroutines in the *same* general-purpose language. We use existing facilities of the language, such as rich libraries, optimizing compilers, and types, to develop concise, declarative, and realistic models with competitive performance on exact and approximate inference. In particular, a wide range of models can be expressed using memoization. Because deterministic parts of models run at full speed, custom inference procedures are trivial to incorporate, and inference procedures can reason about themselves without interpretive overhead. Within this framework, we introduce a new, general algorithm for importance sampling with look-ahead.

## 1 Introduction

Declarative programming is the division of *what* to do and *how* to do it into two modules that can be built and reused separately. In the case of probabilistic inference, the *what* is the definition of a stochastic model, and the *how* is the implementation of an inference algorithm. Dividing the two formally makes it easier to understand and maintain the meaning of the model and the working of the algorithm, especially in complex domains where it is impractical to customize an algorithm to a model by hand coding.

Ever since Bayes nets were first used to represent distributions, declarative programming for probabilistic inference has been studied and practiced extensively (Koller et al. 1997; Getoor and Taskar 2007; Murphy 2007b; Goodman et al. 2008; inter alia). One approach is to encapsulate inference algorithms in a *modeling toolkit*, a library of distributions with operations such as conditionalization, then express models as client programs that invoke the toolkit through its API. Another approach is to express models in a *probabilistic language*, a programming language with constructs such as random choice, then encapsulate inference algorithms in interpreters or compilers for the language.

These two approaches are dual, in that a toolkit operation is run by a model whereas a language implementation runs a model. They also have complementary strengths.

On one hand, the development and use of a modeling toolkit takes advantage of an existing general-purpose language and its facilities such as types, a debugger, and I/O. In particular, if part of a model calls for a custom inference procedure, then the code for the model, written in the same language, can just perform the inference by sidestepping or extending the toolkit. Similarly, if a model needs to refer to an external database, then an existing connection library can be used. On the other hand, the syntax of a probabilistic language can express distributions more succinctly and naturally, as sampling procedures or by generalizing logic programs or relational databases. Also, models in a standalone language may be compiled to more efficient inference code (Fischer and Schumann 2003; Daumé 2007).

This paper presents a new technique for declarative probabilistic inference that combines these strengths. Using a generalization of *coroutines*, we express models as sampling procedures in the *same* general-purpose language in which we implement inference algorithms. Deterministic parts of models are expressed simply as code that makes no random choice, so they run at the full speed of the general-purpose language. Our inference procedures are thus *self-interpreters*, so they can reason about their own accuracy.

### 1.1 A simple example model

We begin to illustrate our monolingual approach using a tiny model (based on Figure 14.11 of Russell and Norvig 2003). To the left in Figure 1 is an influence diagram, in which each node represents a boolean variable. To the right is a corresponding model, expressed as a program in the general-purpose language OCaml (Leroy et al. 2008), for

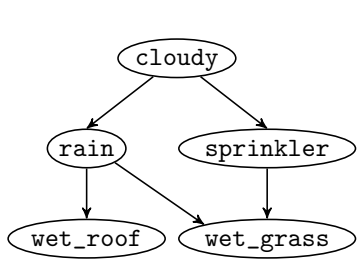

```
let flip = fun p -> dist [(p, true); (1.-.p, false)]

let grass_model = fun () ->
  let cloudy    = flip 0.5 in
  let rain      = flip (if cloudy then 0.8 else 0.2) in
  let sprinkler = flip (if cloudy then 0.1 else 0.5) in
  let wet_roof  = flip 0.7 && rain in
  let wet_grass = flip 0.9 && rain || flip 0.9 && sprinkler in
  if wet_grass then rain else fail ()
```

Figure 1: An influence diagram and a corresponding stochastic model expressed in a *general-purpose* language.

```
let grass_model = fun () ->
  let cloudy    = memo (fun () -> flip 0.5) in
  let rain      = memo (fun () -> flip (if cloudy () then 0.8 else 0.2)) in
  let sprinkler = memo (fun () -> flip (if cloudy () then 0.1 else 0.5)) in
  let wet_roof  = memo (fun () -> flip 0.7 && rain ()) in
  let wet_grass = memo (fun () -> flip 0.9 && rain () || flip 0.9 && sprinkler ()) in
  if wet_grass () then rain () else fail ()
```

Figure 2: Using memoization to improve the performance of the stochastic model in Figure 1.

computing Pr(rain | wet_grass = true). This program uses the function `dist`, which maps a list of probability-value pairs to a randomly chosen value, and the function `fail`, which takes a dummy argument and never returns because it observes an impossible event. Both functions are ordinary OCaml values defined by our framework, not special syntax in a standalone probabilistic language.

For convenience, the code in Figure 1 defines a `flip` function to flip a coin given its probability p of coming up `true`. The code then defines `grass_model`, which takes a dummy argument and either returns a random boolean (namely `rain`) or fails (if `wet_grass` is `false`). These functions are again ordinary OCaml values, of types `float -> bool` and `unit -> bool` respectively. The types `float`, `bool`, and `unit` and their values `0.5`, `true`, and `()` are built-in to OCaml, so a typo such as misspelling `true` is caught by OCaml's type checker at compile time.

Given that `dist` is akin to invoking a random-number generator, and that `fail` is akin to throwing an exception, we can of course perform naïve rejection sampling by applying a higher-order function to the argument `grass_model`. What is less obvious is that more efficient algorithms for exact and approximate inference can also be implemented as higher-order functions that take a model as argument and do not access its source code. The key to these implementations is for `dist` to suspend the execution of the model so that the inference procedure can resume it repeatedly, whether or not from the same point of suspension.

Our model code can use side effects such as mutable state. Thus, we can use *memoization* to express nonparametric models such as Dirichlet processes (Goodman et al. 2008). For this purpose, we provide a higher-order function `memo` (of type `('a->'b)->('a->'b)`), which takes a function as argument and returns a memoized version of it. By applying `memo` mindlessly to *thunks* (that is, functions taking only a dummy argument) or using a simple preprocessor, we can also express *lazy evaluation* to speed up inference. For example, the code in Figure 2 eliminates the moot variable `wet_roof` in Figure 1 (here 'a is `unit` and 'b is `bool`): because OCaml (like most languages) waits until a function is called to evaluate its body, the choice `flip 0.7` in the definition of `wet_roof` is never made, as desired.

### 1.2 The rest of this paper

In §2, we introduce a larger example to show the increased expressivity achieved by placing stochastic models, deterministic computations, and inference procedures all in the same general-purpose language. We analyze performance using the language's I/O facility, then improve performance by invoking inference recursively from within the model.

In §3, we detail the generalized coroutine facility that transfers control between the model and the inference procedure, which lets us reify a model into a tree of choices. This reification enables bucket elimination and, in §4, a new algorithm for importance sampling. We describe our competitive inference performance on realistic models (Jaeger et al. 2007; Pfeffer 2007b). We discuss related work in §5. Our code is at http://okmij.org/ftp/kakuritu/.

## 2 Expressivity

We use Jaeger et al.'s (2007) hidden Markov model (HMM) benchmark to further illustrate how the expressivity of a general-purpose language helps us write clear, fast code.

The HMM is a one-dimensional random walk with 8 states:

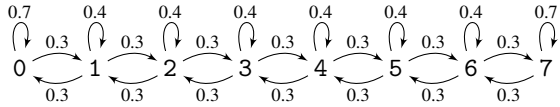

The initial state is chosen uniformly at random. There are 2 observation symbols, L and R. The typical query is to determine the distribution of states after, say, 10 time steps, given some earlier observations.

### 2.1 Types for knowledge representation

The model is specified by the number of states and the transition and observation probabilities. We represent states as integers and define a data type obs of observations L and R.

```
type state = int    let nstates = 8
type obs = L | R
```

If the states and observations get more complex, they should be represented using a structure such as a tuple or object that better matches the problem. Our random variables can be of any type, whether user-defined like obs or built-in like bool, int, tuples, dictionaries, and even functions. We can express distributions over values of all these types without encoding them as bit-strings or numbers.

The transitions are sparse, so we store their probabilities compactly in an array transition_prob of out-edge lists. For example, the array element transition_prob.(2) is the list [(0.4,2); (0.3,1); (0.3,3)]. We initialize the array by giving a general formula that follows the problem description, without repeating boilerplate literals like 0.4. As for observations, we tabulate in another array l_obs_prob the probability of observing L in each state.

```
let l_obs_prob = [| 1.0; 0.85714; ...; 0.0 |]
```

Two simple stochastic functions then express the transition and observation at each time step:

```
let evolve : state -> state = fun st ->
  dist (transition_prob.(st))
let observe : state -> obs = fun st ->
  let p = l_obs_prob.(st) in
  dist [(p, L); (1.-.p, R)]
```

Given a current state, the function evolve chooses the next state, and the function observe chooses the observation. The type annotations above (: state -> ...) clarify the purpose of each function, but they are optional and can be inferred by OCaml. Type errors, such as confusing the state 0 and the observation L, are caught before inference begins.

### 2.2 Higher-order functions for inference

We represent observed evidence as a function that takes a state and time as arguments and may fail. The function run below runs the model for n steps and returns the final state st, asserting observed evidence along the way. It recursively calls itself to run the first n-1 steps, then calls evolve to choose the state transition at step n.

```
let rec run = fun n evidence ->
  let st = if n = 1 then uniform nstates
           else evolve (run (n-1) evidence) in
  evidence st n; st
```

(The function uniform used above is defined in terms of dist to sample from a discrete uniform distribution.) We can pose the conditional query $\Pr(\text{State}_{10} \mid \text{Obs}_5 = \text{L})$ as a thunk query1 that, when invoked, calls run with an evidence function that fails iff L is not observed at time 5:

```
let query1 = fun () ->
  run 10 (fun st n ->
    if n = 5 && observe st <> L then fail ())
```

To conduct exact inference by enumeration, we invoke a function exact_reify whose implementation is described in §3. Applying exact_reify to query1 computes a table that maps states to (unnormalized) probability weights.

Our code expressing the HMM and the query is easy to write and clearly matches the problem description. It is also flexible. For example, because we represent evidence as a function, it is trivial to change the condition from $\text{Obs}_5 = \text{L}$ to $\text{Obs}_n = \text{L}$ for all $n$ between 5 and 10. It is also trivial to extend the model from 8 states to 64: just change nstates (and l_obs_prob if it is not defined by a general formula). There is no literal matrix to enlarge or state encoding to adjust. In contrast, the same change made models an order of magnitude bigger in the systems studied by Jaeger et al.

### 2.3 I/O and self-interpretation for performance

Another way that expressing models in a general-purpose language makes declarative inference more practical is that performance can be profiled and improved using familiar tools of the language. For example, using OCaml's library function Sys.time, we soon discover that exact inference on run takes time exponential in n using exact_reify. Indeed, exact_reify makes an exponential number of recursive calls to run, as any programmer can easily learn using the OCaml profiler or by adding one line to run to increment a counter on every call. The Sys.time function and the call counter coexist indistinguishably with other functions and integers in the model proper.

The reason for the exponential time can be revealed using the OCaml debugger: exact_reify enumerates all possibilities and sums up their probabilities but does not coalesce repeated *intermediate* results in buckets. The Markov property lets us speed up inference, by expressing bucket elimination (Dechter 1999) in terms of `exact_reify`: we just surround `run (n-1) evidence` in the definition of `run` by `dist (exact_reify (fun () -> ...))`. Here `exact_reify` computes a bucket, then `dist` makes a random choice based on the bucket. On this revised `run`, exact inference using `exact_reify` takes time linear in `n`. This improvement is easily confirmed by measuring timings.

This expression of bucket elimination hinges on writing our model code and inference code in the same language: the inference procedure `exact_reify` needs to work on the revised model `run`, which in turn calls `exact_reify`. The recursion depth is unbounded—it is linear in `n`. Moreover, to improve performance at all, the new, inner calls to `exact_reify` need to run as fast as the original, outermost call. Indeed they do, being deterministic parts of the model. In short, our inference procedures are *self-interpreters* that can apply to themselves without interpretive overhead.

To express bucket elimination in general, we can replace any stochastic function `f` by the one below, without accessing the source code of `dist`, `exact_reify`, or `memo`.

```
let bucket = memo (fun x ->
                exact_reify (fun () -> f x)) in
fun x -> dist (bucket x)
```

This expression should be evaluated—and `memo` called—*before* invoking inference, so that all inference shares the same `bucket`. This exact inference strategy handles Jaeger et al.'s benchmarks (2007) in at most a few seconds each.

**2.4  Reasoning about inference procedures themselves**

The ability for models to invoke a variety of inference procedures is unique to our monolingual approach. It is useful not just for performance but also for reasoning about inference itself, such as inference by and about other agents who use approximate inference procedures of their own.

To illustrate this expressivity, we start with a trivial model: Choose a coin that is either fair or completely biased for `true`, with equal probability. Let $p$ be the probability that flipping the coin yields `true`. What is the probability that $p$ is at least 0.3? It is 1, of course, because $0.5 \geq 0.3$ and $1 \geq 0.3$. In the model code below, the predicate `at_least 0.3 true` compares 0.3 against the probability of `true` in the probability tables computed by `exact_reify coin`.

```
let biased = flip 0.5 in
let coin = fun () -> flip 0.5 || biased in
at_least 0.3 true (exact_reify coin)
```

Because a random choice made by an inference algorithm is expressed with `dist` just like any other random choice, any inference procedure, such as the importance-sampling algorithm in §4, can reason about itself or any other inference procedure. For example, suppose we choose a coin as before, then estimate $p$ by flipping the coin twice and dividing the count of `true` by 2. What is the overall probability that our *estimate* is at least 0.3? It is 7/8, because for us to estimate below 0.3 is to choose a fair coin then get `false` from it twice. We can compute this answer in our system by changing `exact_reify` above to a call to sampling twice. This change meaningfully affects what probability tables are tested by `at_least 0.3 true`. As with Goodman et al.'s nested `query` (2008), the outer and inner models may each invoke `fail` to express observations at different levels. However, nested `query` only lets models nest models, not inference, and returns samples, not tables.

This meta-reasoning capability is compatible with memoization. For example, the random variable `biased` above is defined at the outer level of reasoning but used at the inner level (within `coin`), yet we can memoize `biased` as usual:

```
let biased = memo (fun () -> flip 0.5) in
let coin = fun () -> flip 0.5 || biased () in
at_least 0.3 true (exact_reify coin)
```

## 3  Reifying a model into a search tree

As illustrated above, we express models, including any observed evidence, as sampling procedures that may fail. Even without inspecting the source code of models so expressed, rejection sampling is easy: just define `dist` to make a random choice and `fail` to throw an exception. In this section, we explain how to implement `dist` and `fail` differently to support more efficient inference. The bottom line is that we can convert a model to a search tree of random choices. Traversing this tree differently gives rise to exact enumeration (and bucket elimination, as shown in §2.3) and importance sampling (as explained in §4).

To support more than rejection sampling, `dist` should account for multiple possible outcomes of a random choice, not just commit to one of them. For example, calling `flip` should not actually flip a coin but rather explore both outcomes. One way to achieve such nondeterminism is to split the computation into two *threads*, one for each outcome, and merge their results when they finish. This splitting is exactly what the POSIX system call `fork` does, namely to clone the current process to form a new child process. Each call to `fork` returns twice: in the parent process, it returns the child process ID; in the child process, it returns 0.

Because `dist` represents probabilistic choice, we have to track the probability mass of the thread, in a thread-local variable `prob`. For example, evaluating `flip 0.9` should in turn invoke `fork`, then in one thread multiply `prob` by 0.9 and return `true`, and in the other thread multiply `prob` by 0.1 and return `false`. In OCaml syntax, this implementation of `dist` looks as follows.

```
let dist = fun [(p1,v1); (p2,v2)] ->
  if fork () then prob := prob *. p1; v1
             else prob := prob *. p2; v2
```

As for `fail`, it can be implemented as `abort` in POSIX, which terminates the thread and never returns any value.

An inference procedure such as `exact_reify` receives a model as just a thunk. To run the given model, the inference procedure invokes the thunk in a new thread in which `prob` is initialized to 1. Deterministic parts of the model run at the full speed of the general-purpose language, because they invoke neither `dist` nor `fail`. By the time the model finishes, it may finish in any number of threads. For exact enumeration, `exact_reify` can simply accumulate in a probability table the final outcome and probability mass reported by each thread as it finishes.

This implementation strategy works in any language that supports POSIX processes or user-level threads. Most languages qualify, including C, Perl, and Scheme. However, it is inadequate for two reasons, which we address below. Our final implementation requires no OS support for processes and threads; we mention `fork` only for exposition.

### 3.1 Exploring random choices lazily

As the model keeps making random choices, the threads may proliferate so much that running them all in parallel becomes impractical or causes thrashing. Besides, we may only want to run some possible threads. We need tighter control over which possible outcomes to explore: `dist` should suspend the current thread before calling `fork`, and resume only when the inference procedure says to.

More specifically, the suspended thread should provide the inference procedure with a list of possible outcomes and their probabilities, so that the inference procedure can pick which outcomes to explore. Each time the inference procedure requests to explore an outcome, the suspended thread should `fork` off a thread in which `dist` proceeds to return that outcome to the model after updating `prob`. In short, `dist` should turn the current thread into a server to the inference procedure for exploring possible outcomes.

With the implementation of `dist` so revised, an inference procedure receives a model as an *open possibility* in the following sense. An open possibility is a request that can be made to a suspended server thread and yields a *response*. A response is a list of probability-possibility pairs. A possibility is either *closed*, in which case it is just the final outcome value produced by a finished thread, or open. (This recursive definition can be formalized in OCaml as follows.

```
type 'a vc = V of 'a | C of (unit -> 'a pV)
and  'a pV = (float * 'a vc) list
```

We represent a request as a thunk that returns a response.

The type `'a` above is that of final outcomes. The type `'a vc` is that of possibilities (the variant V for closed and the variant C for open), and the type `'a pV` is that of responses.)

Making `dist` suspend the computation thus enables an inference procedure to step through a stochastic model from choice to choice. Still, the fact that OCaml programs are compiled and cannot inspect their own source code or trace their own execution ensures that the deterministic parts of the model run at full speed. That is, we implement `dist` so as to convert, or *reify*, a model to a *lazily* computed tree of choices. Leaves of the tree are closed possibilities, whereas uncomputed branch of the tree are open possibilities. To request an open possibility is to compute the branch.

This tree is a search tree because an inference procedure should search it for leaves with high probability and avoid leaves that `fail`. The model, computing the branches, and the inference procedure, managing the exploration, transfer control to each other like coroutines. For example, `exact_reify` performs depth-first search and accumulates leaves in a probability table as it finds them. This strategy suffices for all the examples so far, but other strategies such as iterative deepening can be used. In §4, we introduce a stochastic search strategy that amounts to importance sampling. It is also possible to represent independent choices compactly using AND nodes in the search tree (McAllester et al. 2008); the use of self-interpretation to express bucket elimination in §2.3 can be viewed as such a representation.

### 3.2 Generalizing coroutines to lightweight threads

POSIX processes and `fork` are rather heavyweight facilities to use for probabilistic programming, and not all operating systems provide them. In comparison, user-level threads can be much more efficient; for example, Erlang programs routinely create millions of simultaneous threads. User-level threads also ease storage management (unused threads can be garbage-collected) and obviate marshalling final outcomes from the model to the inference procedure.

Following Filinski (1994), our OCaml implementation uses a library (Kiselyov 2006) of *delimited control operators* (Felleisen et al. 1987; Danvy and Filinski 1990), which generalize coroutines and user-level threads. (Analogues of the library are available for Haskell, Scheme, and some SML implementations.) The library offers two operations on the execution stack:

- *Reset* pushes a *delimiter* onto the stack, as if installing an exception handler.
- *Shift* pops frames off the stack until it encounters a delimiter, as if throwing an exception, but it captures the frames and creates a function that can be called to push them back onto the stack. The frames constitute a *delimited continuation*, which is like a Common Lisp restart but can be reinstated multiple times.

Reset plays the role of creating a new thread. Our implementation of reification invokes reset to delimit each model from the rest of the program. The delimiter stays on the stack while the model runs using younger frames. If the model finishes with a final outcome, then the delimiter is removed like an unused exception handler, and the outcome is returned to the rest of the program as a closed possibility.

Shift plays the role of suspending a thread and turning it into a server. Our implementation of dist invokes shift to transfer control from the model to the rest of the program, whose stack frames are beyond the delimiter. The delimited continuation captured by shift can be used by the rest of the program to resume the model's execution. Our implementation of fail also invokes shift, but discards the captured continuation, so it is exactly like throwing an exception.

Besides dist and fail, we also provide a higher-order function memo for memoization, used in §1.1 and §2.3. The implementation of memo is straightforward, except each thread (that is, each open possibility) must maintain its own table of memoized results. For example, the repeated calls to the thunk rain in Figure 2, like the repeated references to the variable rain in Figure 1, should always return the same result in each thread, but that result may be true in one thread and false in another. Thus, memo must not mutate the global heap, but rather use or emulate storage local to each thread in which memo is called (Haynes 1987).

## 4 Importance sampling with look-ahead

Given a model reified as a search tree, rejection sampling is just traversing the tree from the root to a leaf, using the probabilities specified at each branch to make a random choice. If this traversal is so lucky as to avoid failure, then the leaf it reaches can be reported—that is, accumulated in a histogram—as a final outcome with weight 1.

In many realistic models, the observed evidence is very improbable, so rejection sampling takes too long to yield enough samples. Importance sampling (Fung and Chang 1990; Shachter and Peot 1990) is a well-known improvement. To perform importance sampling on probabilistic programs written in IBAL, Pfeffer (2007b) developed several sophisticated techniques, which amount to call-by-need evaluation and pushing evidence towards choices.

Pfeffer's techniques require analyzing the source code of the model. Our inference procedures do not have that luxury, because OCaml programs cannot inspect their own source code. Still, we can adapt the idea to operate on a reified search tree: at each branch, we look ahead briefly into each child to sniff out any shallow failures, so as to choose a random child using tighter upper bounds on each child's probability mass. This adaptation has the performance advantage that models can be compiled by OCaml for speed and their deterministic parts run at full speed.

Given a model (as an open possibility, as described in §3.1), our importance-sampling algorithm proceeds as follows.

1. Initialize the weight $p_c$ to 1.
2. Request the given model to get $R$, a response, that is, a list of probability-possibility pairs.
3. Loop until $R$ is empty:
   If $R$ consists of a single element $(p, P)$ where $P$ is an open possibility, then set $p_c$ to $p_c \cdot p$, request $P$ to get a new response $R$, and continue with the loop.
   Otherwise, initialize $L$ to the empty list of probability-response pairs, then for every element $(p, P)$ of $R$:
   (a) If $P$ is a closed possibility, then report the final outcome $P$ with the weight $p_c \cdot p$.
   (b) Otherwise, request the open possibility $P$ and call the response $R'$. If $R'$ consists of a single element $(p', v)$ where $v$ is a closed possibility, then report the final outcome $v$ with the weight $p_c \cdot p \cdot p'$. Otherwise, let $p_t$ be the total probability in the list $R'$, and add the pair $(p \cdot p_t, R')$ to the list $L$.

   Let $p_t$ be the total probability in the list $L$. Quit if $p_t$ is zero. Otherwise: Randomly choose an element $(p, R')$ from $L$ with probability proportional to $p$. Set $p_c$ to $p_c \cdot p_t$. Set $R$ to the result of normalizing $R'$. Continue.

This algorithm may report multiple fractional outcomes on each run, because it treats shallow successes like shallow failures and only stops when $R$ is empty. It subsumes likelihood weighting for Bayes nets, because a choice observed right away to be equal to some value is set to that value.

### 4.1 Data structures with stochastic components

For the look-ahead in our importance-sampling algorithm to help, each random choice needs to be observed soon after it is made, and unobserved random choices should not be made at all. Therefore, our models should be coded using lazy evaluation, or Pfeffer's *delayed evaluation* (2007b). As demonstrated in §1.1, lazy evaluation can be expressed in terms of memoization. Moreover, a composite data structure such as a tuple or list should subject each of its components separately to lazy evaluation, so that, for instance, two lists of independent coin flips can be appended without actually determining any of the flips. To this end, we eschew OCaml's built-in list type and define our own type of *lazy lists*, whose components are memoized thunks:

```
type 'a llist = unit -> 'a lcons
and  'a lcons = LNil
  | LCons of (unit -> 'a) * 'a llist
```

We can append two lazy lists yet not compute any element:

```
let rec lappend y z = memo (fun () ->
  match y () with LNil -> z ()
  | LCons (h,t) -> LCons (h, lappend t z))
```

| Motif pair | 1 | 2 | 3 | 4 | 5 | 6 | 7 |
|---|---|---|---|---|---|---|---|
| Source length | 10 | 9 | 6 | 16 | 18 | 10 | 6 |
| Destination length | 6 | 9 | 6 | 13 | 18 | 10 | 5 |
| IBAL accuracy | .93 | 1 | .28 | .80 | .98 | 1 | .63 |
| 90 sec accuracy | .98 | 1 | .29 | .87 | .94 | 1 | .77 |
| 30 sec accuracy | .92 | .99 | .25 | .46 | .72 | .95 | .61 |

Table 1: The lengths of the 14 motives in the music model and the estimated accuracies of Pfeffer's and our samplers.

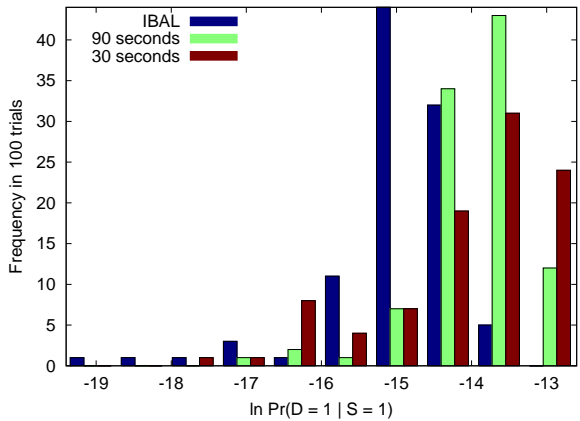

Figure 3: A typical histogram of log-likelihoods produced by Pfeffer's and our samplers on the music model.

### 4.2 Inference performance on realistic models

To gauge the efficacy of this approximate inference strategy on problems of realistic complexity, we reimplemented the classical-music model that Pfeffer (2007b) built to test his importance sampler. The model is of motivic development in early Beethoven piano sonatas. A motif is a list of notes. To develop a motif is to randomly and recursively divide it into a binary tree, then randomly and recursively delete or transpose each subtree. The leaves of the resulting tree form a new motif (using `lappend` above). The randomness and recursion make the number of possible developments exponential in the length of the source motif.

We ran our importance sampler as well as Pfeffer's IBAL sampler on 49 inference problems, each to compute the likelihood that a given source motif $S \in \{1, \ldots, 7\}$ develops into a given destination motif $D \in \{1, \ldots, 7\}$. The goal is to infer the maximum-likelihood $S$ from $D$. The ground truth is that $S = i$ iff $D = i$, but the two motives are different lists of notes—Table 1 shows their lengths. Exact inference is already infeasible for length 5, and the likelihoods are on the order of $10^{-7}$, so rejection sampling is hopeless.

For each problem, we ran 100 trials in which Pfeffer's sampler had 30 seconds per trial (following his original testing), 100 trials in which our sampler had 30 seconds per trial, and another 100 trials in which our sampler had 90 seconds per trial. Using these trials, we estimated the probabilities that the samplers choose the correct $S$ for each $D$. Table 1 shows these accuracy estimates, which suggest that our sampler is competitive with Pfeffer's.

Focusing on a typical inference problem, Figure 3 shows a histogram of $\ln \Pr(D = 1 \mid S = 1)$ sampled. This plot excludes one IBAL trial and five 30-second trials that returned likelihood 0. On this problem, IBAL's likelihood estimates had mean $\exp(-14.7)$ and standard deviation $\exp(-15.3)$; our 90-second sampler's had mean $\exp(-13.8)$ and standard deviation $\exp(-14.5)$; and our 30-second sampler's had mean $\exp(-13.7)$ and standard deviation $\exp(-14.0)$.

Besides the music model, we reimplemented Milch et al.'s model of radar blips for aircraft tracking (2007). In this model, a $10 \times 10$ radar screen monitors a region in the air with an unknown number of planes that move and turn randomly. At each time step, each plane causes a blip on the screen with a certain probability. Due to limited resolution, several planes may result in a single blip. Blips may also be caused by noise. The problem is to estimate the number of planes from blips and their absence in consecutive radar screenshots. A further step is to identify the planes.

To keep inference tractable, we subjected each location coordinate of a plane or blip separately to lazy evaluation, and we used nested inference (as explained in §2.3) to turn our importance sampler into a particle filter. Our model takes many of its parameters from David Poole's AILog 2 code.

When we set the noise probability low, the exact solution could be computed and matched our sampling results. With more noise, our sampling results stayed reasonable. For example, we let there be up to 7 planes (a geometric distribution with ratio .15), detected with probability .9, and up to 4 noise blips (geometric with ratio .02). Having seen the blips $(3,5), (3,7), (3,9)$ at time 0, our sampler found any number of planes possible between 1 and 6, but 3 was most likely, with conditional probability .835. After further observing the blips $(4,5), (4,7), (4,9)$ at time 1 and $(5,5), (5,7), (5,9)$ at time 2, there remained only the possibilities of 3 planes (with conditional probability .987) and 4.

## 5 Related work

Our work is distinguished and motivated by the newfound expressivity and performance afforded by writing models and inference procedures in the same language, especially for deterministic code. All of our code is written in OCaml. In contrast, previous modeling toolkits and probabilistic languages are not implemented in languages they handle. That is, they are not self-interpreters. For example:

- The Bayes Net Toolbox (Murphy 2007a) is a MATLAB library, but its models are not expressed in MATLAB using `rand`, so it cannot reason about itself.

- IBAL (Pfeffer 2007a) is implemented in OCaml, but it cannot reason about OCaml code as is, such as itself.
- Church (Goodman et al. 2008) and Probabilistic Scheme (Radul 2007) are both based on Scheme and implemented using Scheme with mutable state, but they cannot reason about their own implementations, such as about their own inference accuracy.
- Infer.NET (Minka et al. 2009) is implemented on top of C# and compiles models represented in a language that resembles C#, but it cannot reason about C# code as is, such as itself.
- AutoBayes (Fischer and Schumann 2003) and HBC (Daumé 2007) are compilers of specialized languages for statistical models, not of the languages that they target (such as C) or are implemented in (Prolog and Haskell), so they cannot reason about themselves.

Self-interpretation is in principle trivial to achieve, say by writing an interpreter of OCaml in IBAL, but it would be impractically slower than running OCaml code directly, by a factor exponential in the number of interpretation levels. In contrast, our approach incurs no interpretive overhead.

We build on functional rather than logic programming, because we find a functional language more natural for expressing procedural knowledge such as inference algorithms. This choice sets our monolingual approach apart from PRISM (Sato 2008), BLOG (Milch et al. 2007), AILog (Poole and Mackworth 2009), and Markov logic (Domingos and Richardson 2007).

To conclude, we have presented monolingual probabilistic programming, which lets one write declarative stochastic models, deterministic computations, and inference procedures all in the same language. We use mature implementations such as OCaml (SML and Haskell are easily possible). Our approach can express a broad range of models concisely and is amenable to efficient inference. The ability to reify a stochastic program into a lazy search tree lets users write optimized inference procedures, such as importance sampling with look-ahead. The optimizations let us handle realistic models with competitive performance.

### Acknowledgments

We thank Olivier Danvy, Noah D. Goodman, Michael L. Littman, Vikash K. Mansinghka, Avi Pfeffer, Daniel Roy, Stuart Russell, and Matthew Stone for discussions.